\documentclass[a4paper,11pt]{article}
\usepackage{jcappub} 

\usepackage{aas_macros}
\usepackage{wrapfig}
\usepackage{nicematrix}
\usepackage{array}

\usepackage{float}

\usepackage{amssymb}
\usepackage{amsmath} 
\usepackage{color,units}
\usepackage{xspace}
\usepackage{subcaption}
\usepackage{comment}
\usepackage{xcolor}
\usepackage{bm}
\usepackage{dcolumn}
\usepackage{enumitem} 
\usepackage{array}
\usepackage{nicematrix}
\usepackage{hhline}
\usepackage{graphicx}
\usepackage{multirow}
\usepackage[normalem]{ulem}
\usepackage{soul}
\usepackage[math]{cellspace}
\usepackage{nicematrix} 
\usepackage{makecell}
\usepackage{acronym}
\usepackage[font={small}]{caption}
\usepackage{subcaption}
\usepackage{tikz}
\usepackage{xspace}

\usepackage{hyperref}

\newcommand{\sig}{\mathcal{S}}
\newcommand{\bkg}{\mathcal{B}}
\newcommand{\model}{\mathcal{M}}

\title{Model-independent dark matter detection with the Cherenkov Telescope Array Observatory}

\author[a]{Liam Pinchbeck}
\emailAdd{Liam.Pinchbeck@monash.edu}
\author[a]{Csaba Balazs}
\author[a,b]{Eric Thrane}

\affiliation[a]{School of Physics and Astronomy, Monash University, VIC 3800, Australia}
\affiliation[b]{OzGrav: The ARC Centre of Excellence for Gravitational-Wave Discovery, Clayton, VIC 3800, Australia}

\date{\today}

\begin{document}

\abstract{
Searches for annihilating dark matter are often designed with a specific dark matter candidate in mind.
However, the space of potential dark matter models is vast, which raises the question: how can we search for dark matter without making strong assumptions about unknown physics.
We present a model-independent approach for measuring dark matter annihilation ratios and branching fractions with $\gamma$-ray event data. 
By parameterizing the annihilation ratios for seven different channels, we obviate the need to search for a specific dark matter candidate.
To demonstrate our approach, we analyse simulated data using the \texttt{GammaBayes} pipeline.
Given a 5$\sigma$ signal, we reconstruct the annihilation ratios for five dominant channels to within 95\% credibility. 
This allows us to reconstruct dark matter annihilation/decay channels without presuming any particular model, thus offering a model-independent approach to indirect dark matter searches in $\gamma$-ray astronomy. This approach shows that for masses between \unit[0.3-2.5]{TeV} we can probe values below the thermal relic velocity annihilation weighted cross-section allowing a 2$\sigma$ detection for 525 hours of simulated observation data by the Cherenkov Telescope Array Observatory of the Galactic Centre.
}

\maketitle

\section{Introduction}

For the past century, observations of galactic rotation curves, gravitational lensing, large-scale structure formation and more provide strong evidence for the existence of dark matter \cite{Bertone_2018, Zwicky1933gu, 2009_Zwicky_Translation}. However, in spite of the unparalleled success of the standard model of particle physics, it does not explain the nature of dark matter.
Indirect dark matter searches, which look for standard model by-products of dark matter decay or annihilation, provide a promising avenue for discovery of dark matter particles (see e.g. \cite{Slatyer_2022} for a review or \cite{Arcadi2017} for a broader overview on dark matter search methods).
These searches typically focus on observing areas that would likely host a large amount of dark matter, such as the Galactic Centre.
If the annihilation cross section is sufficiently large, these reactions produce a significant $\gamma$-ray flux which does not get deflected by magnetic fields between the source and the detector \cite{Funk_Review_2014, Prandini_2022}. 
The main drawback of trying to use $\gamma$-rays is the large amount of non-trivial conventional astrophysical backgrounds.

Further interest in this search is motivated by the construction of the first telescope in the next generation of imaging atmospheric Cherenkov telescope arrays, the Cherenkov Telescope Array Observatory (CTAO).
Previous literature has shown that CTAO's Galactic Plane survey can produce 95\% exclusion curves for continuum dark matter signals reaching below the dark matter thermal relic annihilation cross-section \cite{CTA2021}.
A persistent problem with indirect searches for dark matter annihilation is that the sensitivity depends on the specific particle physics model assumed for the search and the DM distribution in the targeted environment. In many studies, for example, it is assumed that dark matter will annihilate or decay into a single final state such as $W^+ W-$ or $\tau \bar{\tau}$ (for a review on these techniques we refer again to \cite{Slatyer_2022} and the references therein).
In this paper we propose a model-independent approach that avoids the need to choose a particular dark matter model, while considering multiple standard model final states. 
By taking this approach we can exclude dark matter models by whether they can produce ratios within the constraints produced. 
We implement this framework using a nested mixture model. 

The paper is organised as follows. In Section \ref{sec:PhenModelPhysics} we provide detail into the physics that underpin the various dark matter signal components and how it compares to the literature. Section \ref{sec:mixture_modelling} we detail the mixture model used within this approach and considerations that must be taken. In Section \ref{sec:Demonstration} we provide a demonstration of the kind of information one would obtain with a 5$\sigma$ detection result when taking this approach. We present our projected sensitivities as a function of dark matter mass in Section \ref{sec:ProjSigmaV} and give our final conclusions in Section \ref{sec:Conclusion}.

\section{Indirect Dark Matter Gamma-Ray Spectra}
\label{sec:PhenModelPhysics}

\begin{figure}
    \centering\includegraphics[width=0.8\textwidth]{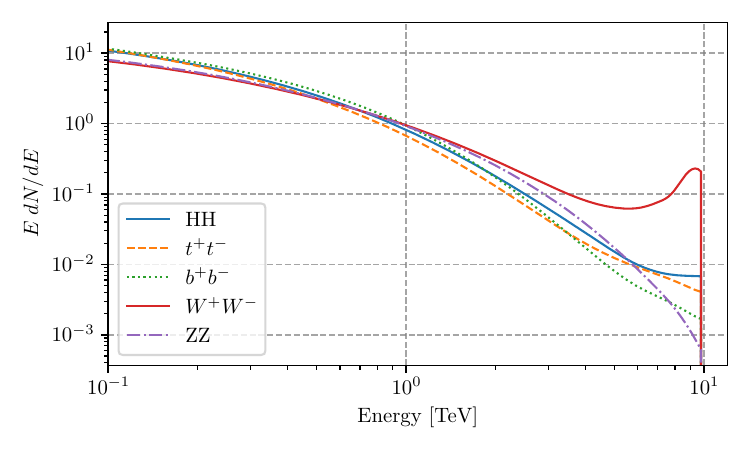}
    \caption{The $\gamma$-ray spectra for dark matter annihilation with a dark matter mass of $\unit[10]{TeV}$ for the standard model final states of $W^+W^-$ (solid, red), $ZZ$ (dash dotted, purple), $HH$ (solid, blue), $t^+t^-$ (dashed, orange), and $b^+b^-$ (dotted, green).
    }
    \label{fig:ChannelSpectra}
\end{figure}

For most indirect dark matter searches with a continuum energy spectrum (as well as line emission searches e.g. \cite{CTAO_Line_Emission}) it is presumed that dark matter particles annihilate or decay into standard model particles. This produces observable signals such as charged cosmic rays (e.g. \cite{DM_CCR_ref1, DM_CCR_ref2}), neutrinos (e.g. \cite{DM_neutrino_ref_1, DM_neutrino_ref_2}) and importantly for this paper, $\gamma$-rays (e.g. \cite{Abdalla_2022, MAGIC_Fermi_2016_paper}). Each type of standard particle, or annihilation/decay `channel', has its own characteristic $\gamma$-ray spectrum for a given dark matter mass. 
We have reliable predictions for the spectra of these final standard-model states. 
We utilise the tabular results contained within the Poor Particle Physicists Cookbook for indirect dark matter detection with electroweak corrections \cite{PPPC_1, PPPC_2}. 
Some examples of these spectra are shown in Fig. \ref{fig:ChannelSpectra}.

The total $\gamma$-ray spectrum produced by the dark matter annihilation is a weighted sum of each spectra $dN/dE$. The weights of this sum are the annihilation ratios $B_f$ of the cross-sections (or branching fractions in the case of decay) for dark matter converting through these channels $\sigma_{f\bar{f}}$ over the the total annihilation cross section $\sigma_{tot}$, $B_f = \sigma_{f\bar{f}}/\sigma_{tot}$. 
It follows that the annihilation ratios obey the relation
\begin{align}
    \sum_f B_f = 1 .
\end{align}

The different annihilation ratios determine the differential flux of $\gamma$-ray events,
\begin{align}
    \frac{d^2\Phi}{dE d\Omega} = \frac{1}{4\pi} \frac{dJ}{d\Omega} \frac{\langle \sigma v \rangle}{2 S_\chi m_\chi^2} \sum_f B_f \frac{dN}{dE} .
    \label{eq:DM_Differential_Flux}
\end{align}
The velocity-weighted annihilation cross section is denoted $\langle \sigma v \rangle$, the dark matter mass $m_\chi$, and the symmetry factor $S_\chi$ represents whether the dark matter particles are there own antiparticles ($S_\chi=1$) or not ($S_\chi=2$). 
A specific dark matter model is described by a given set of $B_f$ that controls the contributions for the various $\gamma$-ray sub-spectra. 
In order to avoid choosing a single dark matter model, we instead constrain the annihilation ratios themselves.  

The angular dependence of Eq. \ref{eq:DM_Differential_Flux} is contained within the differential $J$ factor, denoted $dJ/d\Omega$, which represents the line of sight integral of the mass density $\rho$ in the case of decay and the mass density squared $\rho^2$ in the case of annihilation (two particles are needed in an annihilation but only one in decay). 
In this work we focus on dark matter annihilation with the Einasto dark matter density profile \cite{EinastoRef, EinastoReviewRef} defining
\begin{align}
    \frac{dJ}{d\Omega} = \int_{\textrm{l.o.s}} \left(\rho_s \exp\left(-\frac{2}{a_E}\left[\left(\frac{r(\ell)}{r_s}\right)^{a_E}-1\right]\right)\right)^2 \; d\ell .
\end{align}

\section{Model-Independent Signal Framework}
\label{sec:mixture_modelling}

In this section we develop a phenomenological framework for the dark matter signal. 
It is ``model-independent'' in the sense that it is sufficiently flexible to describe a large class of dark matter candidates, which annihilate or decay into standard model particles.
In this framework, dark matter annihilates into different ``channels,'' each associated with a different standard model particle.
The annihilation or branching ratio into each channel is implicitly determined by the nature of the dark matter particle.
For example, in our previous paper \cite{pinchbeck2024gammabayes}, these ratios were determined by the fact that we assumed a scalar singlet dark matter model.
Here, however, we treat the ratios as free parameters.
This allows us to simultaneously search for a large class of dark matter particles.
To implement this framework we use a nested mixture model.

\begin{figure}
     \centering
     \begin{subfigure}[b]{0.5\textwidth}
         \centering
         \includegraphics[width=\textwidth]{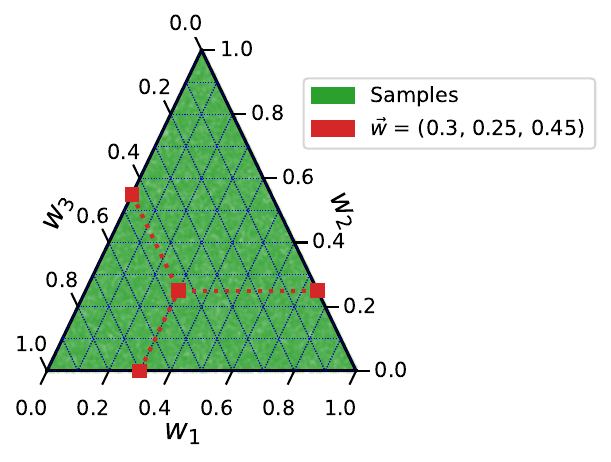}
        \caption{}

     \end{subfigure}
     \hfill
     \begin{subfigure}[b]{0.49\textwidth}
         \centering
         \includegraphics[width=\textwidth]{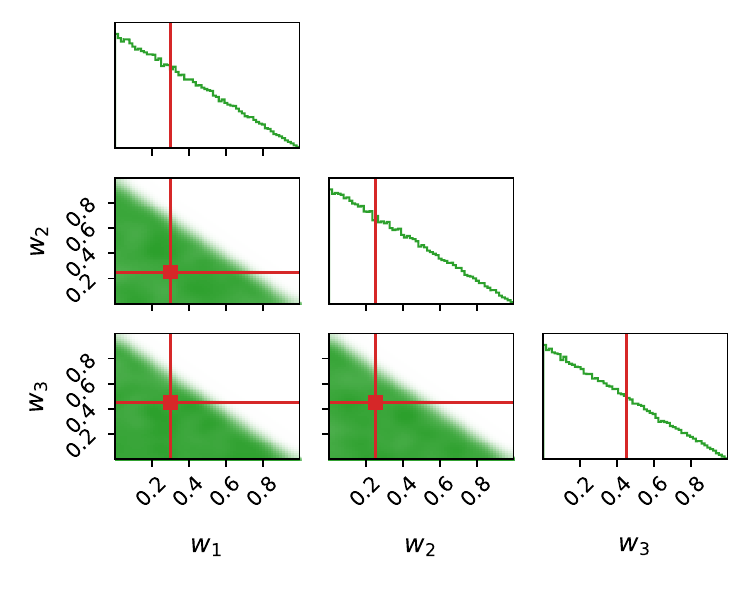}
         \caption{}
     \end{subfigure}
     \hfill
     \begin{subfigure}[b]{0.5\textwidth}
         \centering
         \includegraphics[width=\textwidth]{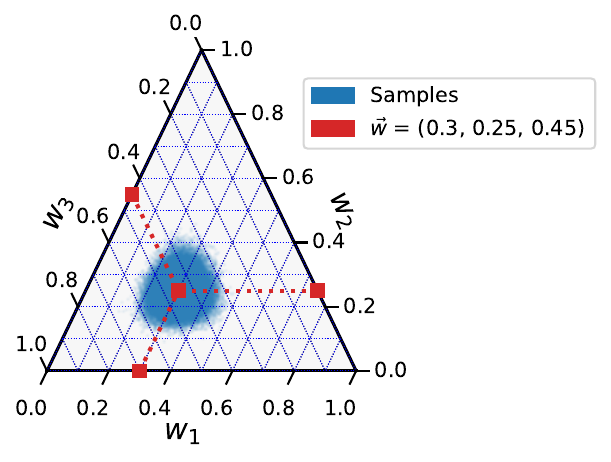}
        \caption{}
     \end{subfigure}
     \hfill
     \begin{subfigure}[b]{0.49\textwidth}
         \centering
         \includegraphics[width=\textwidth]{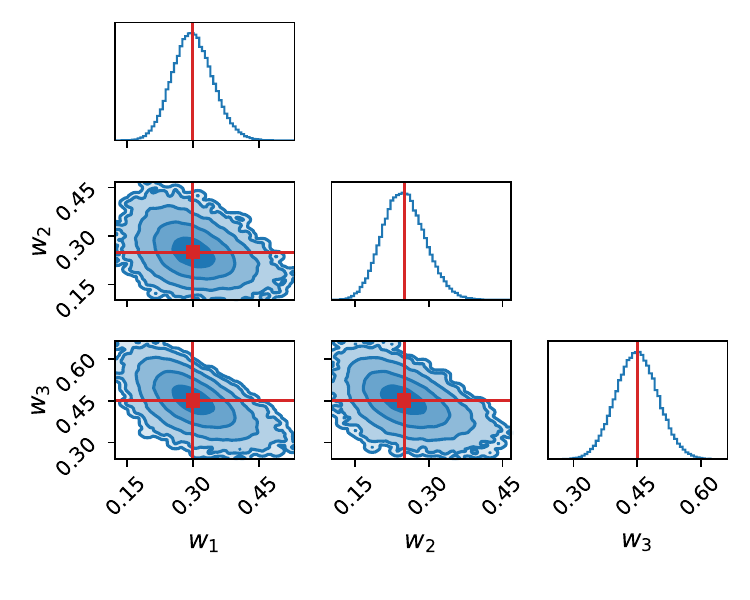}
         \caption{}
     \end{subfigure}
     \caption{
     Prior for the different annihilation fractions $w_i$.
     Plot (a) is a ``ternary plot,'' showing the distribution of three dependent variables subject to the constraint that $w_1+w_2+w_3=1$.
     In this case, the Dirichlet parameters are $\alpha_1=\alpha_2=\alpha_3=1$.
     One point in parameter space is highlighted in red.
     Plot (b) shows the same distribution, but rendered as a corner plot.
     The same point is highlighted in red.
     Plot (c) is a ternary plot similar to (a) except we have changed the $\alpha$ values to $\vec{\alpha}=(30, 25, 45)$.
     Plot (d) is the corresponding corner plot for (c).
     }
     \label{fig:Dirichlet3}
\end{figure}

We adopt a Dirichlet prior for the weights $\vec{w}$
\begin{align}
    \pi(\vec{w}|\vec{\alpha}) = \frac{1}{\beta(\vec{\alpha})} \prod_{i=1}^N w_i^{\alpha_i-1} ,
    \label{eq:Dirichlet_Dist}
\end{align}
where $B(\vec{\alpha})$ represents the beta function of the alpha values, the number of components $N$ satisfies $N\geq2$, and component weights $\vec{w}$ satisfy, 
\begin{align}
    \sum_i^N w_i = 1 .
\end{align}
The Dirichlet distribution is a natural choice of prior for a mixture model with three or more components.
The $\vec\alpha$ hyper-parameters control the shape of the Dirichlet distribution; see Fig.~\ref{fig:Dirichlet3}.

A graphical ``ternary plot'' representation of this distribution in the case of three components are shown in Fig.~\ref{fig:Dirichlet3}. 
The first row of plots shows the Dirichlet distribution when all the alpha values are equal to 1. 
This is a specific example of a symmetric Dirichlet distribution, where all the component alpha values are the same, describing the state of the probabilities on the various weights before receiving any information to constrain them. 
The reason they are seemingly biased towards zero is because if any single weight is large, this implies that the others must be smaller to compensate. 
One cannot have a uniform distribution on all the weights for $N\geq3$ as this would violate unitarity.
The second row demonstrates what the Dirichlet prior can look like when we \textit{do} have prior information on the weights. 
If all of the alpha values are larger than one, the relative ratios of the alpha values is the mode of the distribution with the sum of the alpha values indicating the strength of the prior distribution (see section 22.1 in \cite{Gelman} for more info).\footnote{If one sets one of the alpha values below 1, one can see from Eq.~\ref{eq:Dirichlet_Dist} that the distribution becomes infinite at $w=0$ for this weight. 
This means that the distribution becomes ill-defined in addition to being an improper prior. 
The larger the sum of these values the smaller the variance about this mode.}

\begin{figure}[H]
    \centering
    \includegraphics[width=0.7\textwidth]{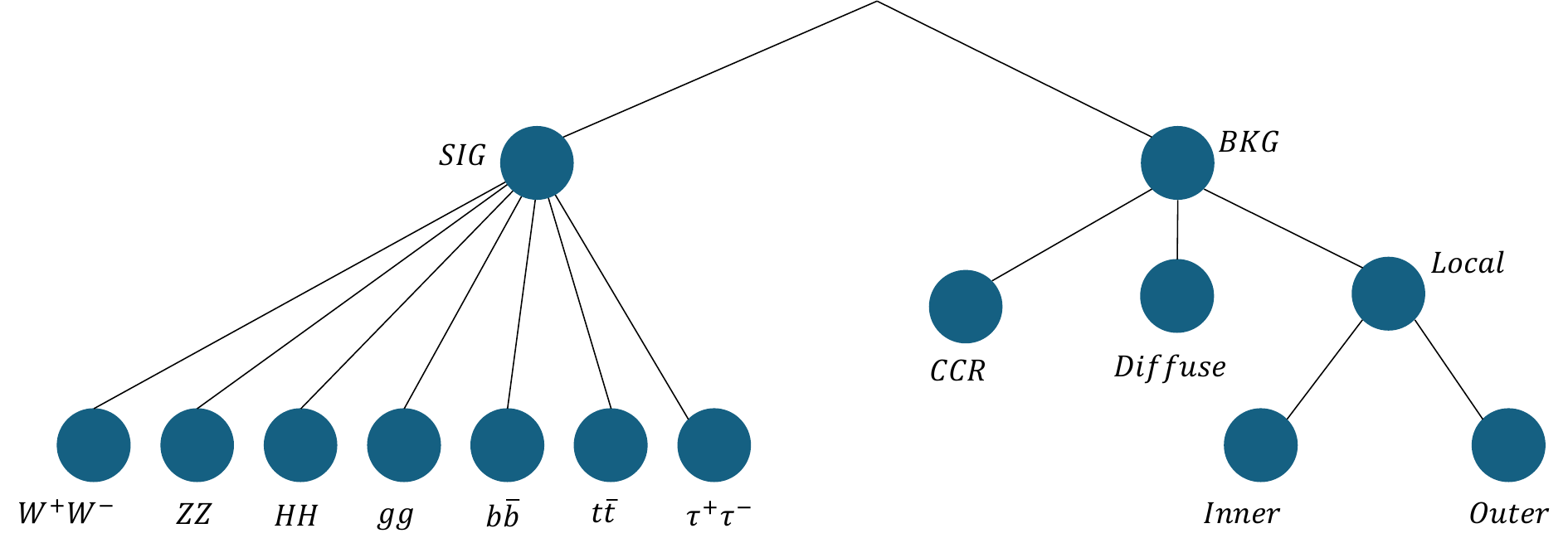}
    \caption{Tree diagram representing the prior on our mixture parameters.
    Each node represents a mixture model with a different associated weight.
    }
    \label{fig:mixture_tree}
\end{figure}

In this work we employ four Dirichlet priors: a \textit{two}-component distribution for the relative fractions of the signal and backgrounds of the total number $\gamma$-ray events, a \textit{three} component distribution for the different background components, a \textit{two}-component distribution for the relative contributions of the localized backgrounds for inner (<1$^\circ$) and outer (>1$^\circ$) galactic radii , and a \textit{seven}-component Dirichlet prior for the different dark matter annihilation ratios. 
One can think of this framework creating a kind of tree structure as shown in Fig. \ref{fig:mixture_tree}, with each node representing the relative fraction of events that can be attributed to the next level of nodes. 
Formally,
\begin{align}
    &\pi(w_\sig, w_\bkg, w_{\bkg i}, w_{\bkg j}^\textrm{local}, B_f | \alpha_\sig, \alpha_\bkg, \alpha_\xi, \alpha_\bkg^\textrm{local}, \alpha_{Bf}) \nonumber \nonumber\\ 
    = & \left(\frac{1}{\text{Beta}(\alpha_\sig , \alpha_\bkg)} w_\sig ^{\alpha_\sig -1} w_\bkg ^{\alpha_\bkg -1}\right)  \left(\frac{1}{\text{Beta}(\alpha_{f})} \prod_{f} B_{f}^{\alpha_{f}-1}\right) \nonumber\\
     &\times\left(\frac{1}{\text{Beta}(\alpha_{\bkg i})} \prod_{i} w_{\bkg i}^{\alpha_{\bkg i}-1}\right) \left(\frac{1}{\text{Beta}(\alpha_{\bkg i}^\textrm{local})} \prod_{j} \left(w_{\bkg j}^\textrm{local}\right)^{\alpha_{\bkg j}^\textrm{local}-1}\right) .
    \label{eq:LiamPrior}
\end{align}
The likelihood is
\begin{align}
    &\mathcal{L}(\vec{d}|w_\sig, w_\bkg, w_{\bkg i}, w_{\bkg j}^\textrm{local}, B_f, \vec{\theta}_\sig, m_\chi) \nonumber\\
    &= \prod_k w_\bkg \mathcal{L}(d_k|\bkg, w_{\bkg i}, w_{\bkg j}^\textrm{local})+ w_\sig \mathcal{L}(d_k|\sig, \vec{\theta}_\sig, B_{f}, m_\chi) \nonumber \\
    &= \prod_k w_\bkg \mathcal{L}(d_k|\bkg, w_{\bkg i}, w_{\bkg j}^\textrm{local}) + w_\sig \, \sum_{f} B_{f} \mathcal{L}(d_k|\sig, \vec{\theta}_\sig, m_\chi) .
    \label{eq:mixture_likelihood}
\end{align}
Where the background likelihood $\mathcal{L}(d_k|\bkg, w_{\bkg i}, w_{\bkg j}^\textrm{local})$ is further described by,
\begin{align}
\mathcal{L}(d_k|\bkg, w_{\bkg i}, w_{\bkg j}^\textrm{local}) =& w_{\bkg \, \textrm{CCR}} \mathcal{L}(d_k|\bkg_\textrm{CCR}) + w_{\bkg \, \textrm{Diffuse}} \mathcal{L}(d_k|\bkg_\textrm{Diffuse}) \nonumber \\ 
&+ w_{\bkg \, \textrm{Local}} \sum_j w_{\bkg j}^\textrm{local}, \mathcal{L}(d_k|\bkg_\textrm{Local}^j).
\end{align}

\begin{figure}[H]
    \centering
     \begin{subfigure}[b]{0.35\textwidth}
         \centering
         \includegraphics[width=\textwidth]{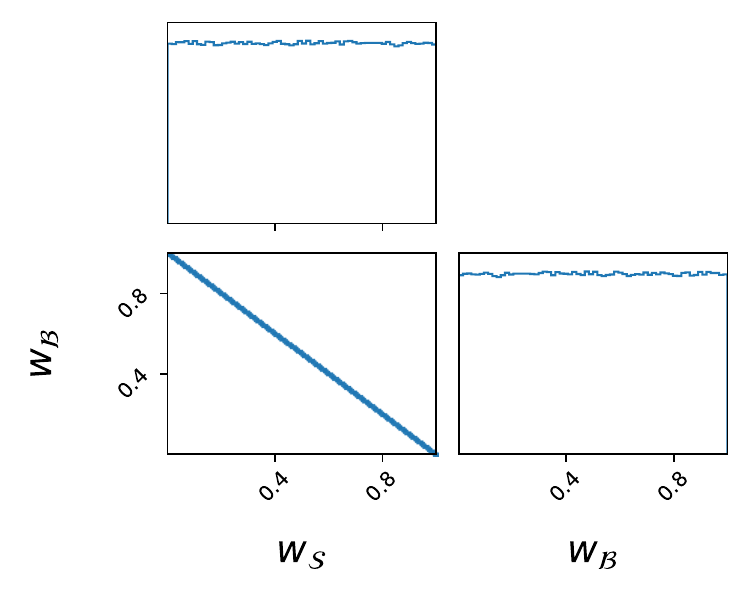}
        \caption{}

     \end{subfigure}
     \begin{subfigure}[b]{0.35\textwidth}
         \centering
         \includegraphics[width=\textwidth]{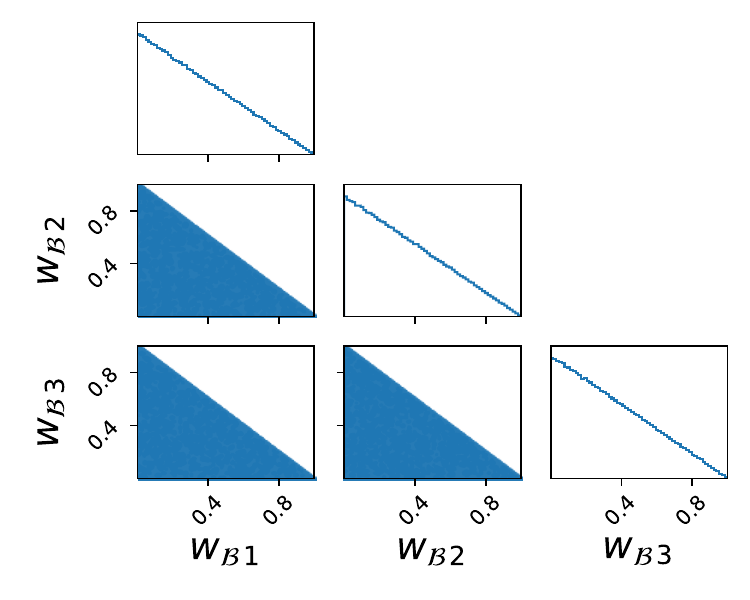}
         \caption{}
        
     \end{subfigure}  
     \begin{subfigure}[b]{0.55\textwidth}
         \centering
         \includegraphics[width=\textwidth]{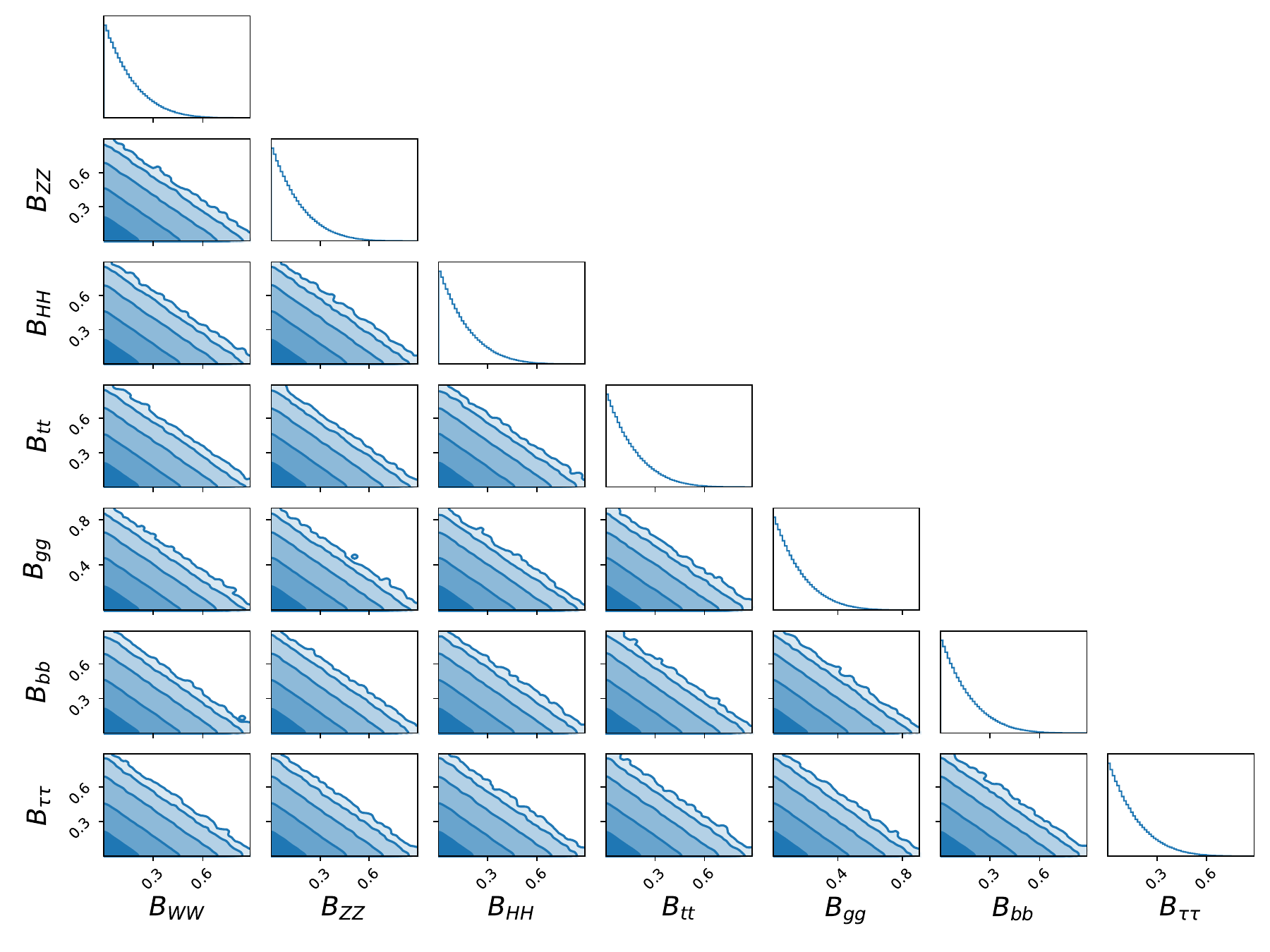}
         \caption{}
   
     \end{subfigure}
    
    \caption{Corner plots containing prior samples from the three Dirichlet priors utilised in this work showing the 1 to 5$\sigma$ credibility contours. Panel (a) shows samples from the two-component Dirichlet prior for the overall signal and background fractions of total events. Panel (b) shows samples from the three-component Dirichlet prior for the relative background fractions of the background components. Panel (c) shows samples from the seven-component Dirichlet prior describing the relative signal component fractions equivalent to the annihilation ratios of the dark matter model.}
    \label{fig:prior_node_samples}
\end{figure}

Here, $ \mathcal{L}(d_k| \vec{\theta})$ is the marginal likelihood of the $k^{\textrm{th}}$ event given the parameters of the event as done in \cite{pinchbeck2024gammabayes}, which we detail in Appendix \ref{sec:NuisanceMarginalisation}. 
Sampling the prior in Eq.~\ref{eq:LiamPrior} leads to the sample distribution shown in Fig.~\ref{fig:prior_node_samples}. 
Fig.~\ref{fig:prior_node_samples} is a corner plot showing the distribution of the overall signal fraction $w_\sig$, background fraction $w_\bkg$, the relative background component fractions $w_{\bkg i}$\footnote{These values are slightly different to the fraction that each respective component contributes to the overall number of events. 
These values are products of the overall weights $w_\sig$ and $w_\bkg$ with the weights of the source components $B_f$ and $w_{\bkg i}$.
The impact of the formulation of our prior in Eq.~\ref{eq:LiamPrior} on these values is discussed in further detail in Appendix.~\ref{sec:ImpliedPrior}.}.
This approach is easily generalised to allow for an arbitary number of signal and/or background components, which, in turn, can split into sub-components.

\begin{figure}[H]
    \centering
    \includegraphics[width=\textwidth]{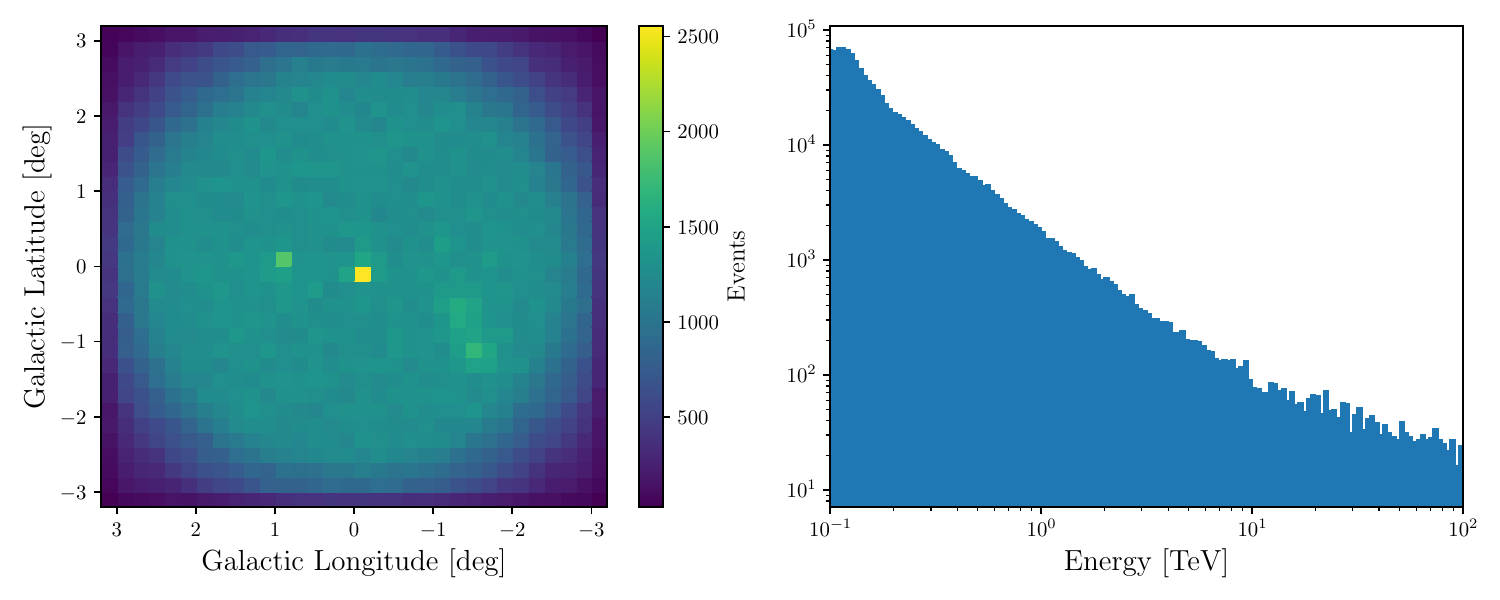}
    \raggedright
    \begin{subfigure}[b]{0.49\textwidth}
    \caption{}
    \end{subfigure}
    \begin{subfigure}[b]{0.5\textwidth}
    \caption{}
    \end{subfigure}
    
    \caption{Plots showing 10$^8$ simulated $\gamma$-ray events of the Galactic Centre observed by CTAO. (a)  Sky map of the angular sky positions of the $\gamma$-ray events. (b) Histogram of the energies of the $\gamma$-ray events over log base 10 scale.}
    \label{fig:DetectionMeasuredEventDistributions}
\end{figure}

\begin{figure}[H]
    \centering
     \begin{subfigure}[b]{0.49\textwidth}
         \centering
         \includegraphics[width=\textwidth]{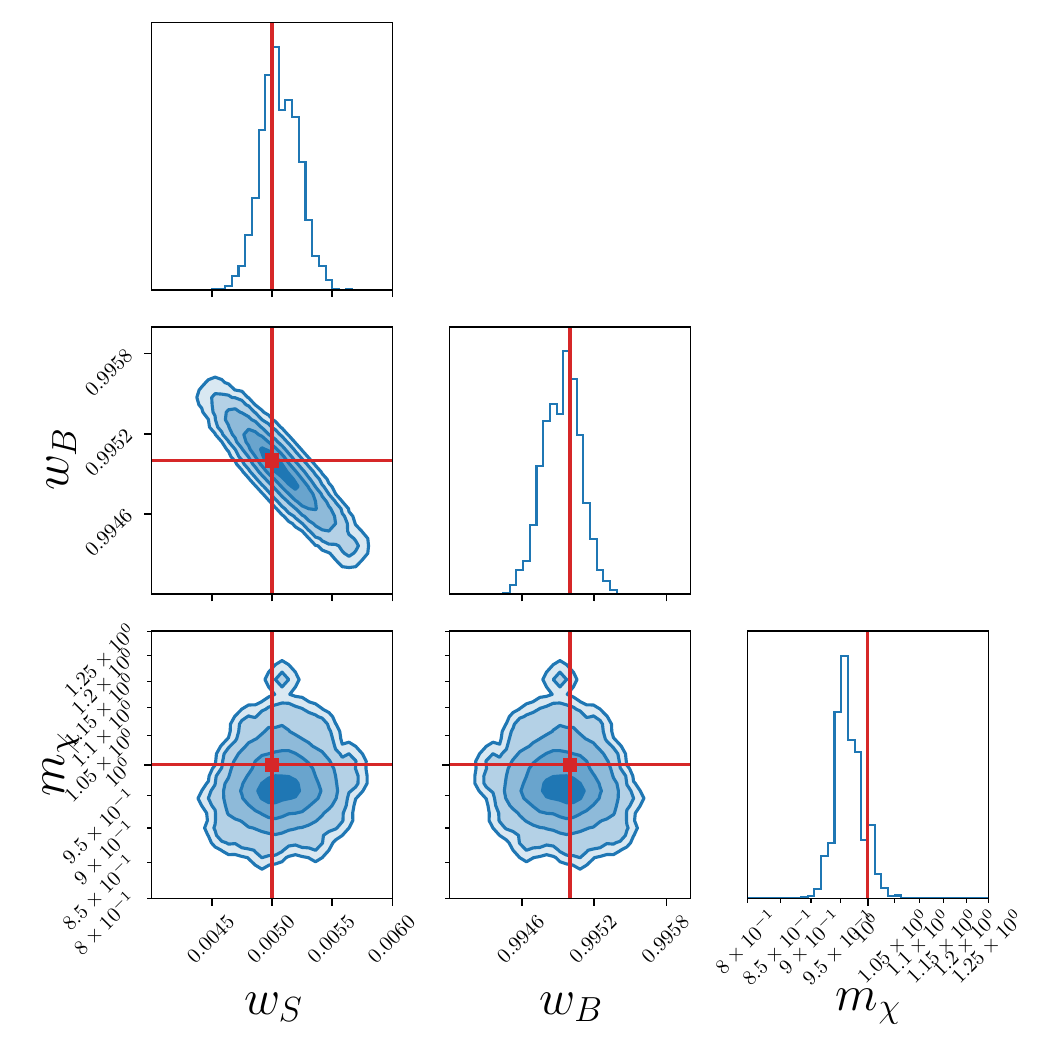}
        \caption{}
        \label{fig:sig_bkg_mass_posterior}

     \end{subfigure}
     \begin{subfigure}[b]{0.49\textwidth}
         \centering
         \includegraphics[width=\textwidth]{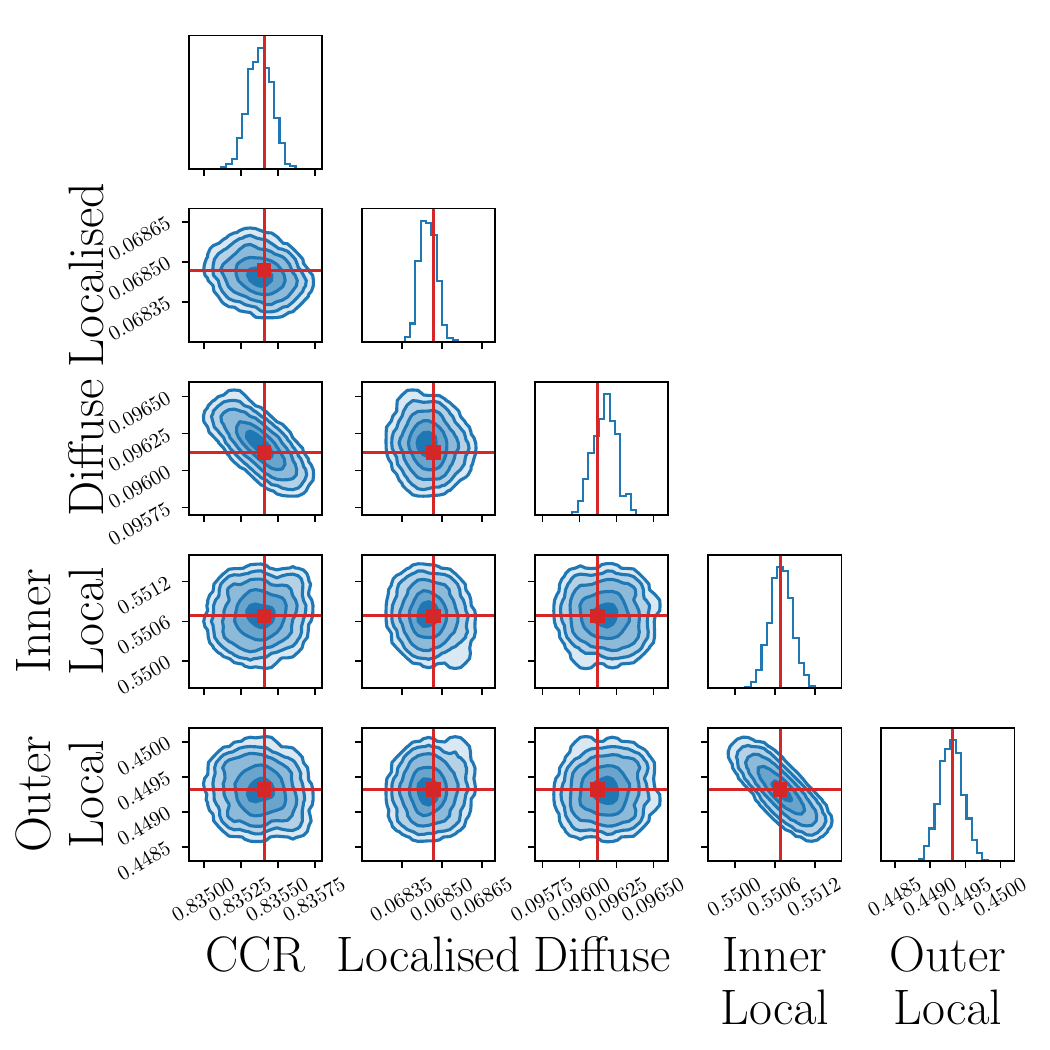}
         \caption{}
        \label{fig:rel_bkg_posterior}
     \end{subfigure}   
     \caption{Corner plots showing the posterior for different mixing fractions and dark matter mass.
     The posterior is obtained using 10$^8$ $\gamma$-ray events from the Galactic Centre which includes 10$^5$ $\gamma$-ray events originating from dark matter annihilation. True values are represented by orange lines and the contours signify the one to 5$\sigma$ credibility levels. As zero is outside the 5$\sigma$ value for the signal fraction, this plot constitutes an example of a 5$\sigma$ detection by CTAO of a dark matter signal.}
    \label{fig:ExDetectionPosterior}
\end{figure}

\section{Demonstration: Recovering dark matter annihilation ratios}\label{sec:Demonstration}

Here we show how one can recover the annihilation ratios from the data without assuming a given dark matter model. 
We simulate $10^8$ events (a rough estimate of the expected number for the CTAO's Galactic Plane survey involving the Galactic Centre \cite{ScienceWithTheCTA}) with angular coordinates within $3^\circ$ of the galactic centre with energy values ranging from \unit[0.1]{TeV} to \unit[100]{TeV}.
We inject 5$\times10^5$ dark matter signal events (equivalent to a signal fraction of 0.005) with annihilation ratios corresponding to the the $\mathcal{Z}_2$ scalar singlet model used in \cite{CTA2021}, except in this analysis we do not assume these ratios when analysing the data; we infer them.
We use a local dark matter density of \unit[0.4]{GeV/cm$^3$} to be compatible within the range of values generally accepted in global fits \cite{local_DM_density_ref}.  
The distribution of simulated true energy and true sky positions are shown in Fig.~\ref{fig:DetectionMeasuredEventDistributions}.


\begin{table}
    \centering
    \begin{tabular}{|c|c|c|}
    \hline
    
   \textbf{Model Parameters}  & \textbf{True Values}\\
    \hline
    \hline
      Signal Fraction & 0.005 \\
      DM Density Model & Einasto \\
      $a_{E}$ & 0.17\\
      Local DM Density & \unit[0.4]{GeV/cm$^3$} \\
      Einasto Scale Radius & \unit[20]{kpc} \\
      DM Model & $\mathbb{Z}_2$ Scalar Singlet\\
      Higgs Coupling Constant ($\lambda_{hS}$) & 0.1\\ 
    \hline
    \hline
    Annihilation Channel & 
    \\
    \hline

    W$^+$W$^-$ & 0.606 \\
    ZZ & 0.308 \\
    HH & 0.062\\ %
    t$^+$t$^-$ & 0.024 \\
\hline
\end{tabular} 
\caption{True values of model parameters used for simulation of 10$^8$ $\gamma$-ray events or roughly 525 hours of observation time of the Galactic Centre by CTAO.  Dashes represent that it would not make sense for the relevant 'parameter' to follow a distribution. Distributions that cover multiple rows indicate that they apply to all the rows they span across. }
\label{tab:Model_Params}
\end{table}

\begin{figure}
    \centering
     \begin{subfigure}[b]{\textwidth}
         \centering
         \includegraphics[width=0.6\textwidth]{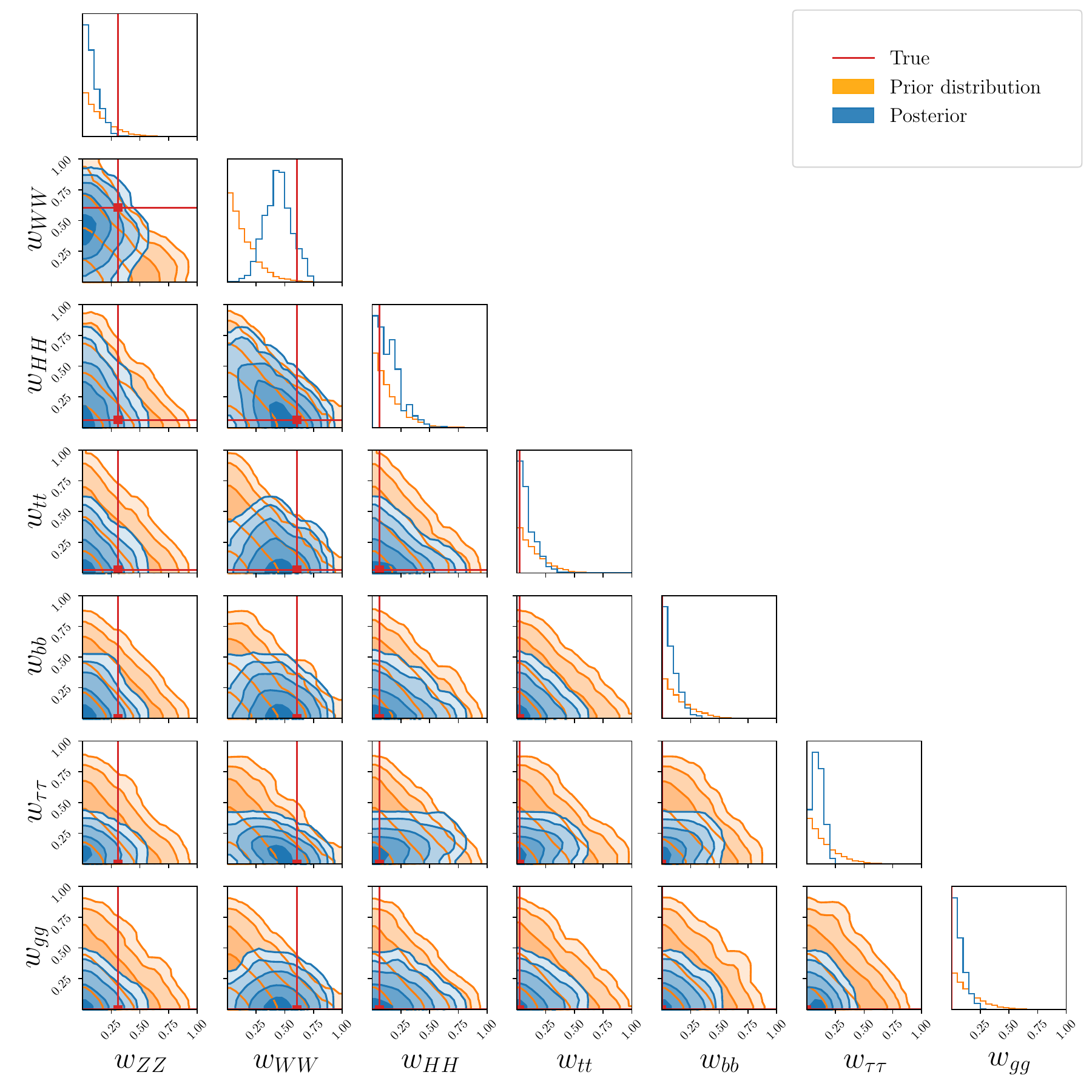}
         \caption{}
     \end{subfigure}  
     \caption{Corner plot showing the posterior of the annihilation ratios in the case of a 5$\sigma$ detection. The $W^+W^-$ channel is localised around 0.48 and excludes a zero contribution with 3$\sigma$ credibility. The other channels cannot be resolved away from zero, though, the $ZZ$ channel can likely be measured with a ten-fold increase in data based on anecdotal testing. These other posterior follow the prior distribution  
     }
    \label{fig:ExDetectionBfPosterior}
\end{figure}

We then use \verb|GammaBayes| (calling \verb|dynesty| \cite{dynesty_ref_1, dynesty_ref_2}) to find the posterior for the model parameters. 
The results of this are shown in Fig.~\ref{fig:ExDetectionPosterior} and Fig.~\ref{fig:ExDetectionBfPosterior}.
Fig.~\ref{fig:ExDetectionPosterior} shows a corner plots of the posterior samples for the signal and background fractions $w_\sig$ and $w_\bkg$ along with the dark matter mass $m_\chi$ in Fig.~\ref{fig:sig_bkg_mass_posterior} and the relative fractions of the charged misidentification, localised source and diffuse backgrounds respectively in Fig.~\ref{fig:rel_bkg_posterior}. We do not show the prior on the localised background sub-components as it is identical to Fig.~\ref{fig:sig_bkg_mass_posterior}.
Fig.~\ref{fig:ExDetectionBfPosterior} shows a corner plot of the posterior samples for the relative signal components or equivalently the dark matter annihilation ratios.
These plots show that we recover the signal at the 5$\sigma$ credibility level as shown in the signal fraction plots in Fig.~\ref{fig:sig_bkg_mass_posterior}, and thus examples of the information one can infer from a 5$\sigma$ detection.

\begin{figure}
    \centering
    \includegraphics[width=0.5\textwidth]{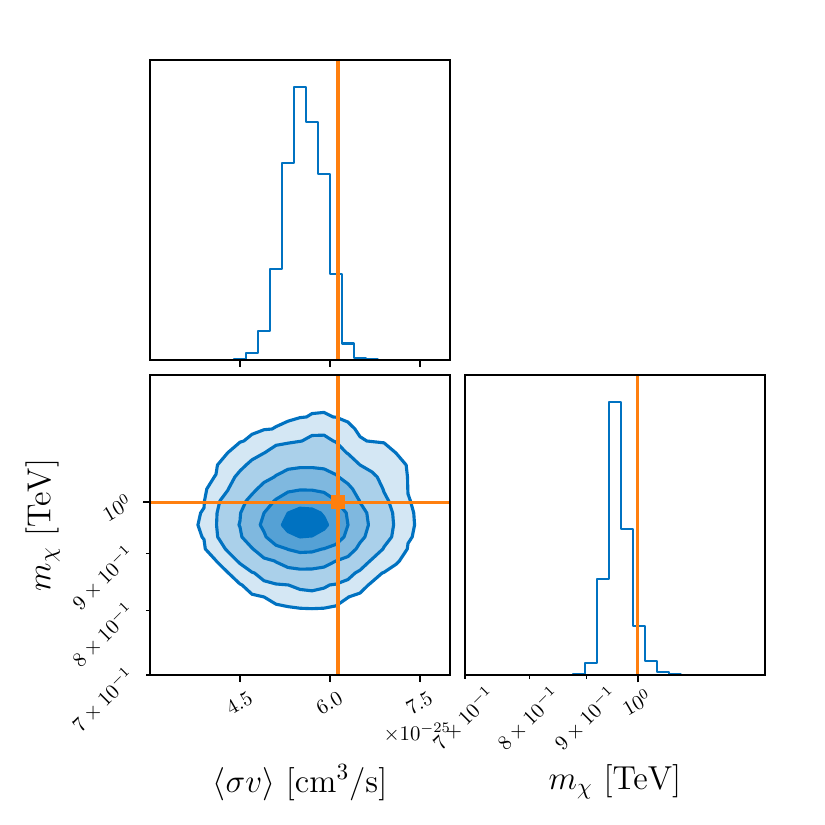}

    \caption{The marginal posterior plots for $\langle \sigma v \rangle$ and the mass $m_\chi$ using the results from Fig.~\ref{fig:ExDetectionPosterior} and Fig.~\ref{fig:ExDetectionBfPosterior}. The true values are represented by the orange lines and are both within the 1$\sigma$ of credibility values. In line with Fig.~\ref{fig:ExDetectionPosterior}, the 5 sigma contour for the distribution of $\langle\sigma v\rangle$ values do not overlap with 0, so it is an example of what a detection result would look like.}
    \label{fig:ExDetectionSigmaV}
\end{figure}

By rearranging Eq.~\ref{eq:DM_Differential_Flux} and using the posterior samples of the mass, signal fraction and annihilation ratios, we produce the posterior on  $\langle \sigma v\rangle$ as shown in Fig.~\ref{fig:ExDetectionSigmaV}.

\section{Projected constraints}
\label{sec:ProjSigmaV}
\begin{figure}
    \centering
    \includegraphics[clip=True, 
    trim={0 0 0 0},
    width=0.7\textwidth]{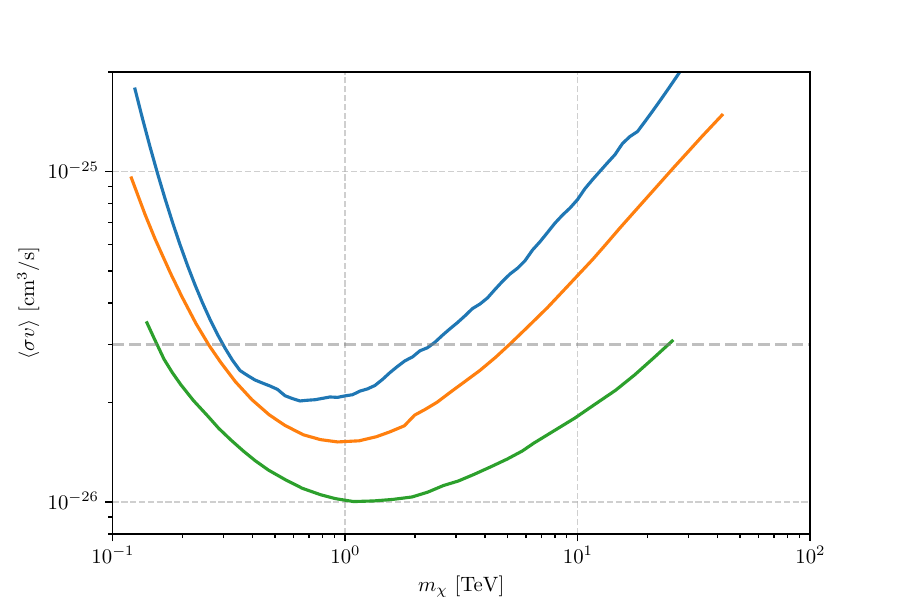}
    \caption{The sensitivity of CTAO to dark matter annihilation using the dark matter model-independent approach described in this paper for 525 hours of Galactic Centre data.
    In blue we plot the expected 95\% credibility upper limits as a function of $m_\chi$ and in orange the upper limits when the relative contributions of the inner and outer local contributions are fixed.
    In green, are the expected frequentist upper limits reproduced from \cite{CTA2021}. 
    The two curves should be compared \textit{qualitatively} since the two analyses employ different background and very different signal models; see the main text for details.
    The horizontal grey dashed line represents the $\langle \sigma v \rangle$ value inferred from cosmological calculations for the current dark matter relic density \cite{Steigman_2012}.}
    \label{fig:Proj_SigmaV}
\end{figure}

In the event of a non-detection, we can place an upper limit on the velocity-weighted annihilation cross-section $\langle \sigma v \rangle$.
We calculate this projected upper limit as a proxy for the sensitivity of this search.
We simulate a new dataset, which is otherwise the same,with the signal fraction set to 0. 
We extract the 2$\sigma$ credibility contour values from the posterior samples of $\langle \sigma v \rangle$ and dark matter mass; see Fig.~\ref{fig:Proj_SigmaV}.
The projected sensitivity is below the thermal relic annihilation cross section shown as the horizontal grey dashed line in Fig.~\ref{fig:Proj_SigmaV}.

\section{Conclusion}
\label{sec:Conclusion}

In this paper we show how to construct a dark matter model independent inference framework for analysing $\gamma$-ray event data. 
This framework allows one to perform inference on the annihilation or branching ratios that quantify the interaction between dark matter and conventional standard model particles. 
With this approach we show that one retains similar levels of sensitivity to previous works \cite{CTA2021, pinchbeck2024gammabayes} while allowing broader classes of dark matter models to be investigated in a single fit. Additionally, in the case of a 5$\sigma$ detection this approach facilitates Bayesian model comparison by whether the annihilation or branching ratios of a model could lie within the constraints calculated by this inference.

Future work will investigate generalising the dark matter density profile and including a parameter for whether the signal comes from annihilation or decay to further reduce the dark matter model independence of approach. 
It may also prove useful to further split the backgrounds into sub-components and investigate different realisations of the diffuse backgrounds to account for systematics in the high energy regime for a more flexible background model. 
Further expansion of this framework will require the implementation of a sample re-weighting scheme or different method for approximating the posterior distribution.
The main issues with larger generalisation currently are the memory usage related to the initial nuisance parameter marginalisation method for multiple prior hyper-parameters.

\appendix

\section{Instrument Response Functions}
\label{sec:IRFs}
In this appendix we provide further detail on the prod5 version 0.1 Instrument Response Functions (IRFs) currently supplied by CTAO \cite{prod5_irfs_ref}. The total observational likelihood is the product of the energy dispersion and point spread functions, which are assumed to be independent. The explicit dependence on the instrument, $\mathcal{I}$, is included here for clarity but is implied for the rest of the paper.

\subsection{Energy Dispersion}

\begin{figure}
     \begin{subfigure}[b]{0.49\textwidth}
         \centering
         \includegraphics[width=\textwidth]{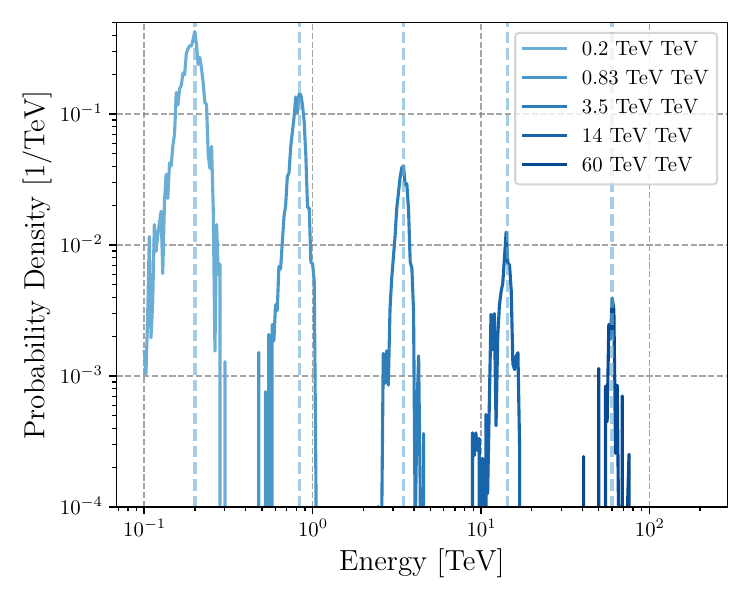}
        \caption{}

     \end{subfigure}
     \hfill
     \begin{subfigure}[b]{0.49\textwidth}
         \centering
         \includegraphics[width=\textwidth]{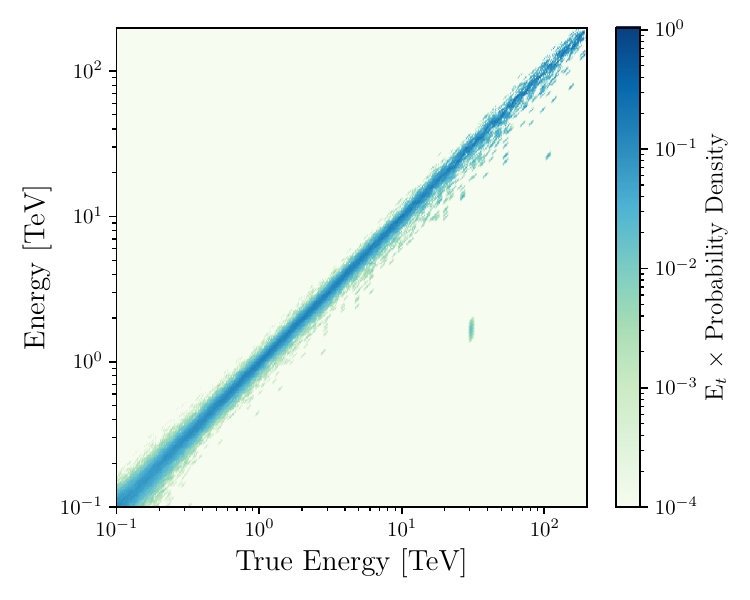}
        \caption{}

     \end{subfigure}
     \hfill
     \caption{Representations of the CTAO's energy dispersion function for a true sky position of of (0$^\circ$, 0$^\circ$) in galactic longitude and latitude. (a) Energy dispersion over reconstructed energy (over which the energy dispersion is normalised) for various true energy values. One can see that the max values of these distributions decrease over time meaning they become less localised in reconstructed energy. (b) Scaled energy dispersion by true energy as a function of reconstructed and true energies. As the true energy increases, the model becomes more noisy.}
     \label{fig:EDISP}
\end{figure}

The energy dispersion, denoted $E_{disp}(E | E_t, \mathbf{\Omega}_t, \mathcal{I})$, is a model for the probability density of a measured energy value $E$ (no subscript) given the true energy and sky position values $E_t, \Omega_t$, normalised as,
\begin{align}
    1 = \int_{E} dE \; E_{disp}(E | E_t, \mathbf{\Omega}_t, \mathcal{I}).
\end{align}
It quantifies the instruments ability to accurately measure the energy of a $\gamma$-ray. For the alpha configuration of CTAO graphical representations of this function are shown in Fig.~\ref{fig:EDISP}.

\subsection{Point Spread Function}

Similar to the energy dispersion, the point spread function, denoted $PSF$, quantifies the instruments' ability to reconstruct the sky position of a given $\gamma$-ray event. Specifically it is the probability density of a reconstructed sky location $\mathbf{\Omega}$ (no subscript) given the true energy and sky position values $E_t$ and $\mathbf{\Omega}_t$ and is normalised such that,
\begin{align}
    1 = \iint_\mathbf{\Omega} d\mathbf{\Omega} \; PSF(\mathbf{\Omega}|E_t, \mathbf{\Omega}_t, \mathcal{I})
\end{align}

\begin{figure}
\centering
     \begin{subfigure}[b]{0.49\textwidth}
         \centering
         \includegraphics[width=\textwidth]{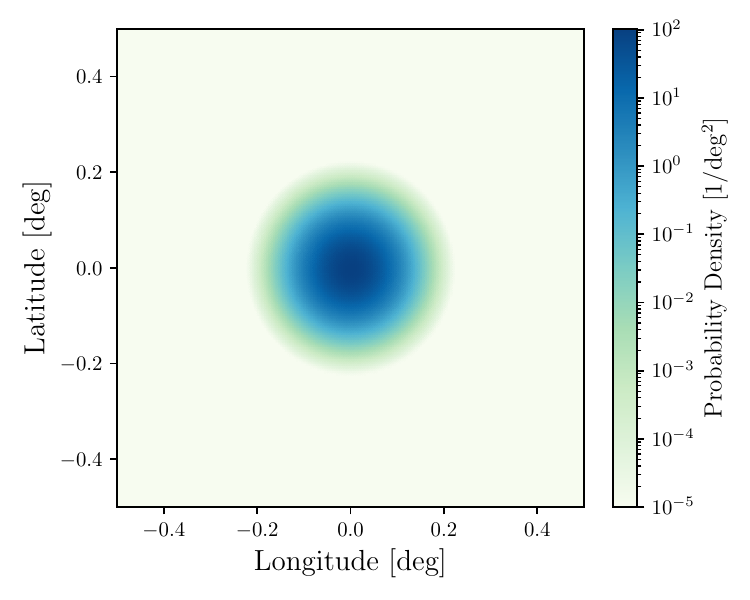}
        \caption{}

     \end{subfigure}
     \hfill
     \begin{subfigure}[b]{0.49\textwidth}
         \centering
         \includegraphics[width=\textwidth]{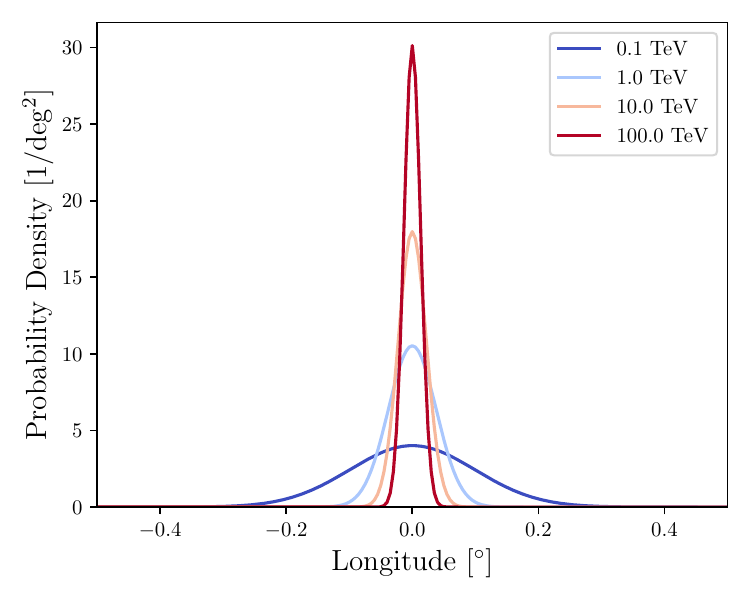}
        \caption{}

     \end{subfigure}
     \hfill
     \caption{Representations of the PSF as supplied by CTAO. (a) 2D representation of the PSF for a true energy value of 1 TeV and true sky position of (0$^\circ$, 0$^\circ$) in galactic longitude and latitude. It is over this space that the PSF is normalised. (b) Slices of the PSF for various true energy values for a fixed reconstructed latitude of 0$^\circ$ and true sky position (0$^\circ$, 0$^\circ$) in galactic longitude and latitude. This shows that as the energy increases CTAO is better at reconstructing the position of a given $\gamma$-ray event.}
    \label{fig:PSF}
\end{figure}

Example representations of the PSF are shown in Fig.~\ref{fig:PSF}.

\subsection{Effective Area}

\begin{figure}
     \begin{subfigure}[b]{0.49\textwidth}
         \centering
         \includegraphics[width=\textwidth]{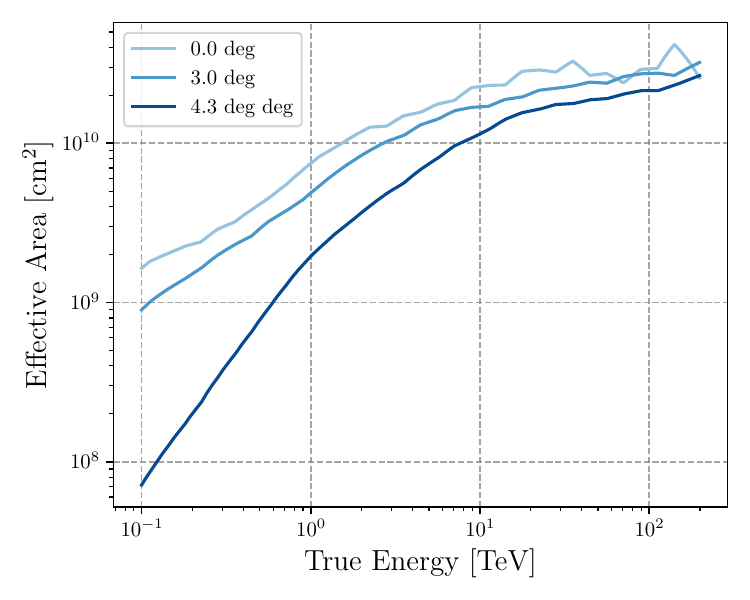}
        \caption{}
     \end{subfigure}
     \hfill
     \begin{subfigure}[b]{0.49\textwidth}
         \centering
         \includegraphics[width=\textwidth]{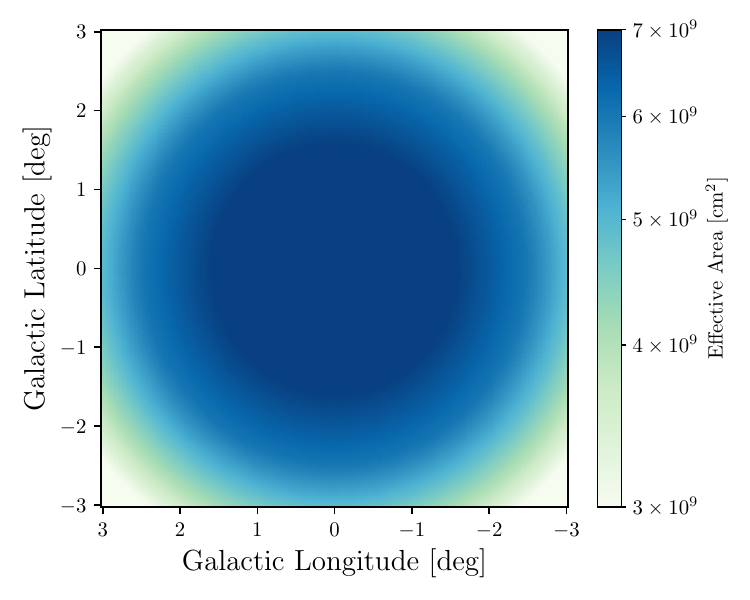}
        \caption{}

     \end{subfigure}
     \hfill
     \caption{Representations of the effective area function produced by CTAO. (a) Slices of the effective area function over true energy, offset by $0^\circ$, $3^\circ$ and $4.3^\circ$ from the pointing axis of the telescope array. The effective area for $\gamma$-ray events decreases for increasing offset and increases for increasing energy. Slice of the effective area for a true energy of 1 TeV. Similar to the other sub-figure, the effective area decreases for increasing offset from the pointing axis. }
    \label{fig:Aeff}
\end{figure}

Unlike the other two IRFs the effective area, $A_{\textrm{eff}}(E_t, \mathbf{\Omega}_t|\mathcal{I})$ does not describe a probability density function. The effective area function gives the effective cross-sectional area of the telescope array to a given set of true $\gamma$-ray event values (units of area) for a given instrument. Some representations of the effective area are shown in Fig~\ref{fig:Aeff}.

Within this framework we utilise observational prior models, $\pi_{\textrm{obs}}(E_t, \psi_t |\mathcal{M}, \mathcal{I})$ (units of counts/energy solid angle time area), which are related to the differential flux models $\pi_{\textrm{flux}}(E_t, \psi_t |\mathcal{M})$ (flux models that contain nothing about about the instrument taking the data with units of counts/energy solid angle time) in the following manner,
\begin{align}
    \pi_{\textrm{obs}}(E_t, \psi_t |\mathcal{M}, \mathcal{I}) \propto A_{\textrm{eff}}(E_t, \psi_t|\mathcal{I}) \pi_{\textrm{flux}}(E_t, \psi_t |\mathcal{M}).
\end{align}
Thus, $\pi_{\textrm{obs}}(E_t, \psi_t |\mathcal{M}, \mathcal{I})$ represents the given probabilities of energy and sky position detected by a given instrument $\mathcal{I}$ (e.g. CTAO) for a given model $\mathcal{M}$ (e.g the signal model $\sig$).

\section{Gamma Ray Backgrounds}
\label{sec:GammaRayBKG}
\begin{figure}
    \centering
    \includegraphics[width=\textwidth]{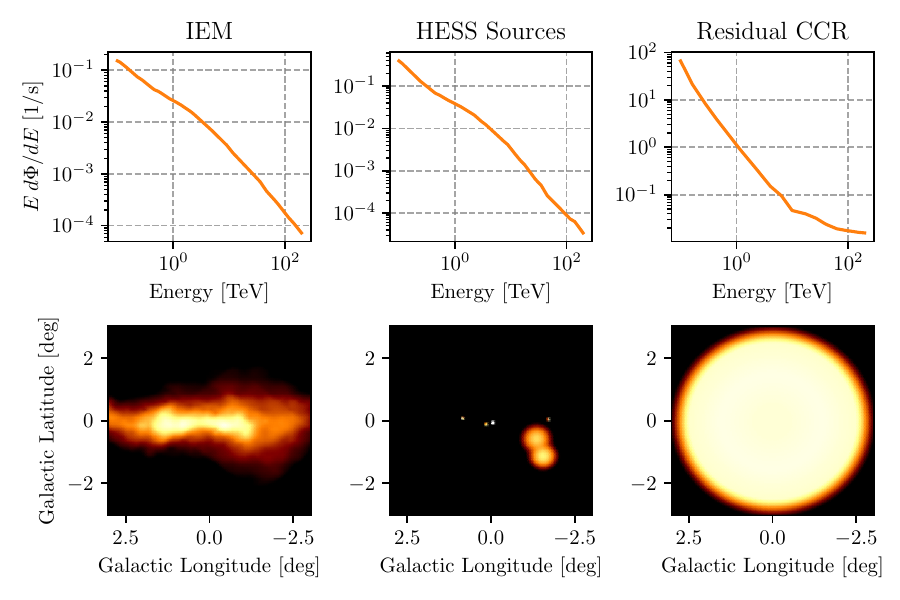}
    \raggedright
     \begin{subfigure}[b]{0.41\textwidth}
        \caption{}
     \end{subfigure}
     \begin{subfigure}[b]{0.22\textwidth}
        \caption{}
     \end{subfigure}
     \vspace{-2.7em}
     \begin{flushright}
     \begin{subfigure}[b]{0.29\textwidth}
        \caption{}
     \end{subfigure}
    \end{flushright}

    \caption{
    Astrophysical background models with the effective area of the telescope applied to the interstellar emission and localised backgrounds. (a) Interstellar emission model with the same morphology as the public \texttt{Pass8} interstellar emission model with the spectrum as in \cite{Gaggero_2017}. 
    (b) Combined event rate from available source flux models in the H.E.S.S catalogue within 3 degrees in Galactic Longitude and Latitude of the Galactic Centre. (c) \texttt{Prod5} Residual Cosmic Ray Background model produced by CTAO corresponding to the alpha configuration of the Southern CTAO site.
    }
    \label{fig:BKG_Models}
\end{figure}

The vast majority of $\gamma$-rays detected CTAO will come from conventional astrophysical backgrounds \cite{Berge_2006_BKG_Modelling, Mohrmann_BKG_Validation}. 
In this paper we categorise these backgrounds into three components: the mis-identification of charged cosmic rays, interstellar emission from around the galactic centre, and $\gamma$-rays that can be attributed to particular sources as contained within the H.E.S.S catalogue. 
These sources are shown in Fig~\ref{fig:BKG_Models}.

\subsection{Charged Cosmic Ray Mis-identification Background}
In any set of $\gamma$-ray ground-based detector observations, the largest background will be the misidentification of $\gamma$-rays or residual charged cosmic rays (e.g. see \cite{Sciascio_2019} and \cite{RowellBook} for a review on analysis methods for ground-based detectors and efforts to minimise this background). 
This can be seen in Fig.~\ref{fig:BKG_Models} with the charged cosmic ray background flux values orders of magnitude higher than the other $\gamma$-ray sources. 
This background is specific to ground-based $\gamma$-ray observatories, which detect $\gamma$-rays from the particle showers produced when they hit the atmosphere (see, e.g. Ref. \cite{weekes_IACT_def_ref} and references therein). 
charged cosmic rays much more commonly produce these showers, and this background reflects the instruments' ability to distinguish these from $\gamma$-ray initiated showers. 
We utilise the charged cosmic ray model that comes along with the set of IRFs (detailed in Appendix \ref{sec:IRFs}) produced CTAO \cite{prod5_irfs_ref}.

We assume that this background is only radially dependent from the centre of the field of the view (FOV) of the given observation.
Different observation runs may have different realisations of this background from MCMC simulations reflecting the conditions of the array during the given observation period \cite{MonteCarloShowerDesignStudies, BernlohrSimTel_2008}. 
The 525 hours of observations that we simulate later on in this paper will come from many observation runs and thus use multiple realisations of this background model. Assuming that the overall behaviour of these realisation is consistent, then we do not expect the this to change the overall results shown here (using a single misidentification model). 
One could carry out a more comprehensive analysis following the approach described in Appendix \ref{sec:NuisanceMarginalisation}.

\subsection{Localised $\gamma$-Ray Backgrounds}
Localised $\gamma$-ray sources as $\gamma$-rays originate from objects such as pulsar wind nebulae, supernova remnants, etc. 
We utilise the H.E.S.S catalogue to extract differential flux models for these sources within $5^\circ$ of the Galactic Centre \cite{HGPS_ref}.\footnote{These sources include J1741-302 (unknown counterpart), J1745-290 (Galactic Centre/unknown counterpart), J1745-303 (Supernova Remnant/molecular cloud), J1746-285 (unknown counterpart), J1746-308 (unknown counterpart), J1747-248 (unknown counterpart), and J1747-281 (Pulsar Wind Nebula) \cite{HGPS_ref}.}
The only source with a confirmed counterpart in this region is J1747-281, which is associated with a pulsar wind nebula.
In our Bayesian framework, $\gamma$-rays with small angular separations from these sources should not contribute much significance to a dark matter signal because they can be explained with the background model. We do, however, further split this background into two components for sources within 1 degree of the Galactic Centre and those outside this boundary. This is to reflect possible systematics or biases that particularly may impact sources around the Galactic Centre due to the complicated nature of that particular region of the sky.

The CTAO is predicted to have a much smaller angular resolution at completion than any previous Imaging Atmospheric Telescope Array array.
Thus, it may identify many new $\gamma$-ray point sources around the Galactic Centre. 
For this reason, some studies include a population of unresolved point sources to see the impact of these on their analysis (see e.g. \cite{CTA2021} and the references therein).
We chose not to do this here as it was outside the scope, but believe it would be interesting to quantify the impact this would have on our analysis.
If this is the case, then the overall results of this study will likely remain unchanged.

\subsection{Interstellar Emission Background}

The interstellar emission background is a diffuse emission component of $\gamma$-rays extending along the Galactic Plane originating from the interaction of cosmic rays with the interstellar gas and radiation fields \cite{Fermi_LAT_IEM_Model}. 
It was the brightest emission component in the Fermi-LAT data of the Galactic Plane \cite{FermiLAT_Diffuse_GR_Obs}. 
We utilise the spatial morphology of the Fermi-LAT \texttt{Pass 8} interstellar emission model and the power-law spectrum developed by \cite{Gaggero_2017}.
This is in an effort to emulate the background model used for this component in \cite{CTA2021} to make later comparison more appropriate. 
We note, however, that the diffuse component, particularly outside of Fermi-LAT's sensitivity range, is not completely understood, and uneven distributions of the Galactic cosmic rays around the Galactic Centre could also potentially bias a result.

\section{Implied mixture weight prior}
\label{sec:ImpliedPrior}
\begin{figure}
    \centering
     \begin{subfigure}[b]{0.4\textwidth}
         \centering
         \includegraphics[width=\textwidth]{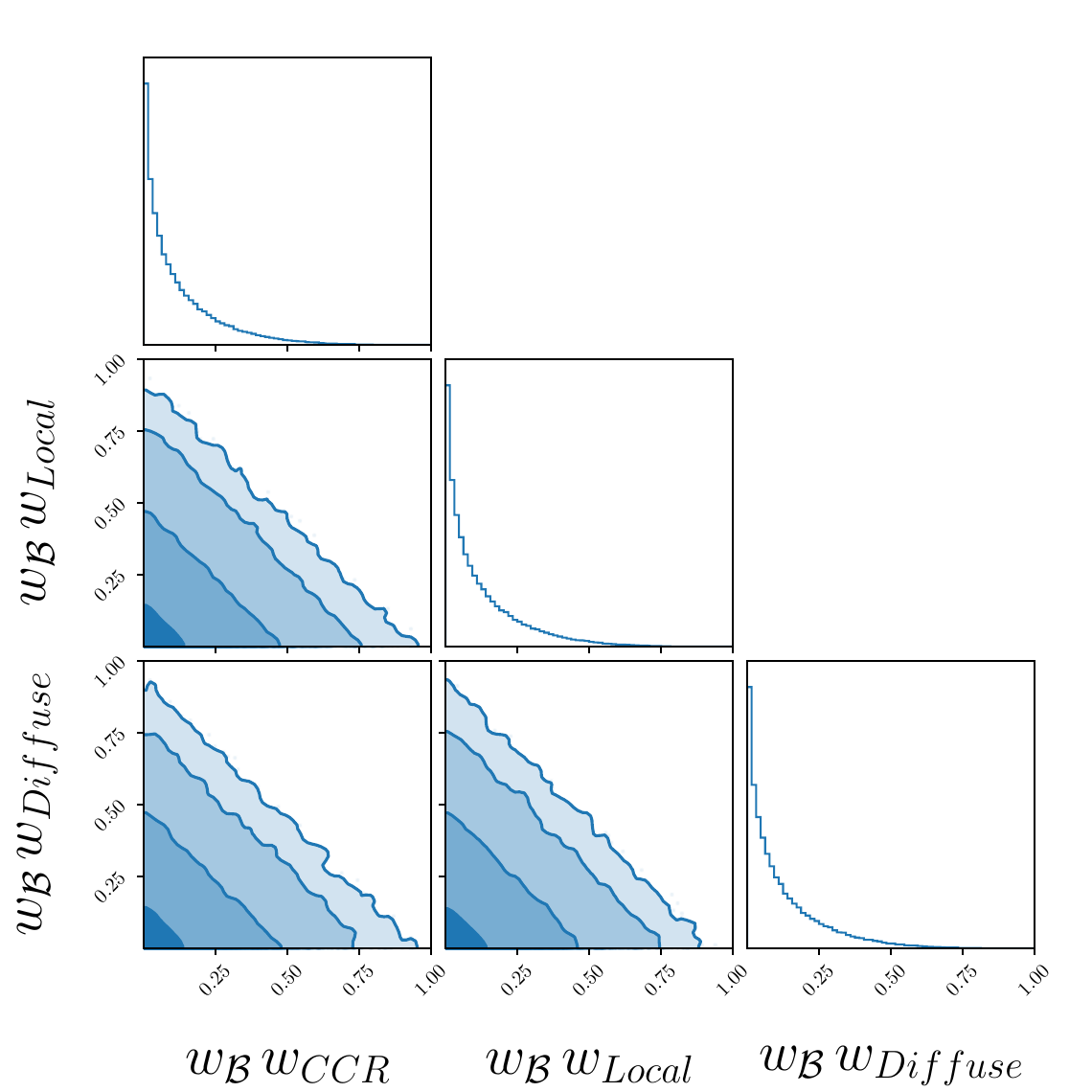}
        \caption{}

     \end{subfigure}
     \begin{subfigure}[b]{0.35\textwidth}
         \centering
         \includegraphics[width=\textwidth]{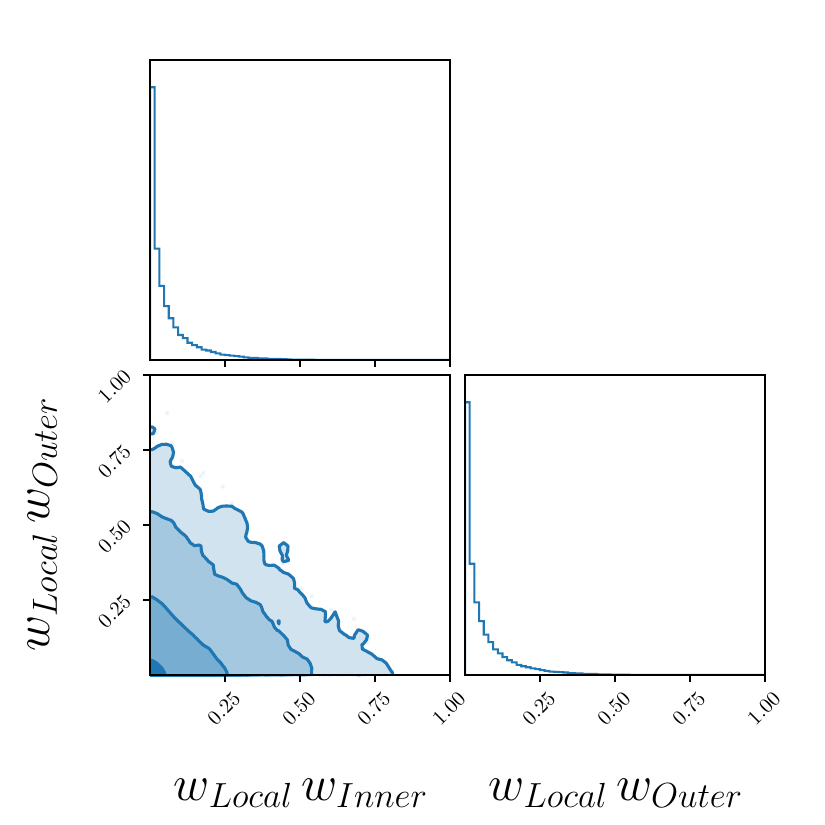}
        \caption{}

     \end{subfigure}
     \begin{subfigure}[b]{0.8\textwidth}
         \centering
         \includegraphics[clip, trim={0.4cm 0 0.3cm 0}, width=\textwidth]{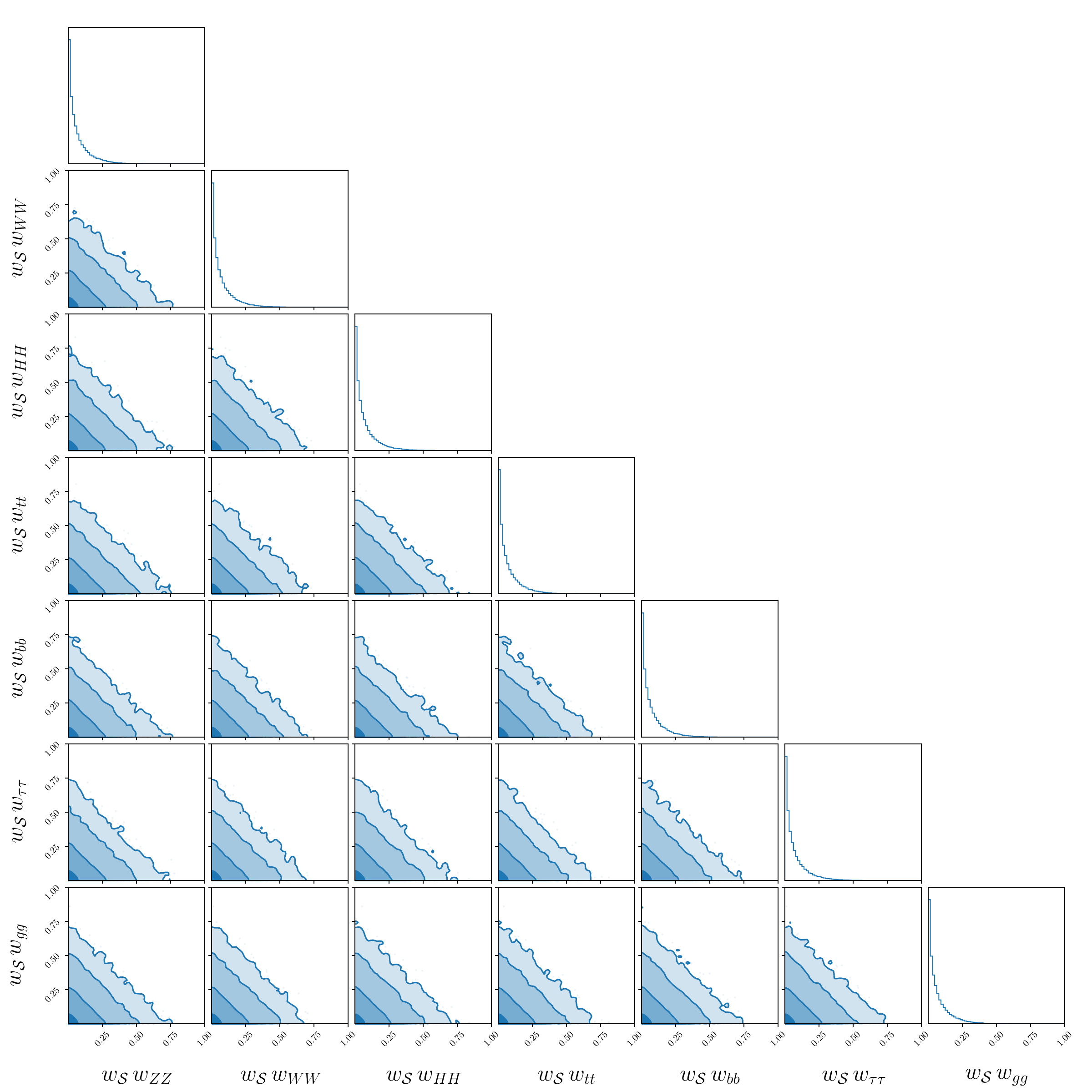}
         \caption{}
    
     \end{subfigure}

    \caption{The implied prior samples for the various mixture weights where the values correspond to how many of the events are coming from the given component. (a) Samples from the three component Dirichlet prior for the relative background fractions of the background components weighted by the overall background fraction. (b) Effective prior samples for the two component inner and outer radii localized source contributions. (c) Samples from the seven component Dirichlet prior describing the relative signal component fractions equivalent to the annihilation ratios of the dark matter model weighted by the overall signal fraction.}

    \label{fig:implied_prior_weight_samples}    
\end{figure}

Fig.~\ref{fig:implied_prior_weight_samples} shows the implied prior distributions on the weights of how much each signal component is assumed to contribute to the total number of $\gamma$-ray events. Each individual dark matter channel component weight ($w_S  B_{f}$) contributes a small fraction of the total. 

\section{Nuisance Parameter Marginalisation}
\label{sec:NuisanceMarginalisation}
In this appendix we detail how the marginalised probabilities in Eq. \ref{eq:mixture_likelihood} are calculated. 
For the $k^{\rm{th}}$ measured event, there are the measured values of said event, $\mathbf{d}^k=(E^k, \mathbf{\Omega}^k)$ (no subscript), and a set of `true' values, denoted $\mathbf{d}_t^k=(E_t^k, \mathbf{\Omega}_t^k)$, which are what one would see if the given instrument were infinitely precise.
Each observational prior (e.g. the background and signal models) is implicitly dependent on these true values, while the likelihood in Eq.~\ref{eq:mixture_likelihood} is dependent on the measured values.
We are not interested in all the individual values of these true parameters, but rather the model parameters dictating the overall distribution of these values. 
Thus, we designate them as nuisance parameters and marginalise over them.

This process of marginalisation is done as shown in Equation \ref{eq:nuisance_marginalisation},
\begin{align}
    \mathcal{L}(\mathbf{d}^k|\vec{\theta_\model}, \model) = & \int d\mathbf{\Omega}^k_t \int dE^k_t \mathcal{L}(\mathbf{d}^k|\mathbf{\Omega}^k_t, E^k_t) \pi(\mathbf{\Omega}^k_t, E^k_t|\vec{\theta_\model}, \model).
    \label{eq:nuisance_marginalisation}
\end{align}
The total observational likelihood evaluated as the normalised product of the energy dispersion and point spread function is denoted $\mathcal{L}(\mathbf{d}^k|\mathbf{\Omega}^k_t, E^k_t)$, the observational prior for a given model $\model$ with relevant model parameters $\vec{\theta}_\model$ is denoted $\pi(\mathbf{\Omega}^k_t, E^k_t|\vec{\theta_\model}, \model)$. This is the most computationally expensive step of the analysis due to the shear number of times it must be computed, as it must be done for every $\gamma$-ray event which number into the hundreds of millions. 

It is thus important that the computation time of this operation is minimised on the per event level. Currently, this is achieved by computing these values discretely, and directly integrating them by utilising batched vector operations. 
For further detail we defer to the documentation for \texttt{GammaBayes} which is being utilised to calculate these values. 

However, we finish on a remark on the computational complexity of this overall Bayesian approach.
The complexity scales linearly if one introduces individual parameters to the observational priors, and exponentially if one adds multiple parameters to the same prior (adding more mixture parameters is sub-linear and less of a concern). 
Meaning, if you have a baseline run with two priors, one with a range of parameter values being tested, and the other with none then this takes $t$ time to complete. 
If we then add a similar range of values for a parameter to the second prior, then the computation time will double to $2t$ time. 
If we instead add another parameter to the \textit{first} prior, then the computation time will be four times what it was previously to $4t$ time. 
Further extension of this approach for priors with many parameters included in the inference will require a sample re-weighting approach, computing the integrals with a nested sampling or MCMC method, that is currently sub-optimal for the specific investigation of this paper.

\acknowledgments{This work was conducted in the context of the CTAO Dark Matter Working Group and was performed on the OzSTAR national facility at Swinburne University of Technology. The OzSTAR program receives funding in part from the Astronomy National Collaborative Research Infrastructure Strategy (NCRIS) allocation provided by the Australian Government, and from the Victorian Higher Education State Investment Fund (VHESIF) provided by the Victorian Government. This research has made use of the CTAO instrument response functions provided by the CTAO Consortium, see https://www.ctao-observatory.org/science/cta-performance/ (version prod5 v0.1; \cite{prod5_irfs_ref}) for more details.
E.T. is supported by ARC CE170100004, LE210100002, DP230103088, and CE230100016.
The research of C.B. is supported by ARC DP210101636, DP220100643, and LE210100015.}

\bibliography{refs}
\bibliographystyle{JHEP}

\end{document}